\documentclass[
 reprint,
 superscriptaddress,
 amsmath,amssymb,
 aps,
 pra,
 floatfix,
]{revtex4-2}
\pdfoutput=1

\usepackage{graphicx}
\usepackage{dcolumn}
\usepackage{bm}
\usepackage[colorlinks=true, linkcolor=black, urlcolor=black, citecolor=black]{hyperref}

\newcommand{\dcwidth}{140mm}
\newcommand{\scwidth}{70mm}

\usepackage[normalem]{ulem}

\begin{document}

\preprint{APS/123-QED}

\title{Generation of a Single-Cycle Acoustic Pulse:\protect\\A Scalable Solution for Transport in Single-Electron Circuits}

\author{Junliang Wang}
    \altaffiliation{These authors contributed equally to this work.}
    \affiliation{Univ. Grenoble Alpes, CNRS, Grenoble INP, Institut N\'eel, 38000 Grenoble, France}
\author{Shunsuke Ota}
    \altaffiliation{These authors contributed equally to this work.}
    \affiliation{Department of Electrical and Electronic Engineering, Tokyo Institute of Technology, Tokyo 152-8550, Japan}
    \affiliation{National Institute of Advanced Industrial Science and Technology (AIST), National Metrology Institute of Japan (NMIJ), 1-1-1 Umezono, Tsukuba, Ibaraki 305-8563, Japan}
\author{Hermann Edlbauer}
    \altaffiliation{These authors contributed equally to this work.}
    \affiliation{Univ. Grenoble Alpes, CNRS, Grenoble INP, Institut N\'eel, 38000 Grenoble, France}
\author{Baptiste Jadot}
    \affiliation{Univ. Grenoble Alpes, CNRS, Grenoble INP, Institut N\'eel, 38000 Grenoble, France}
    \affiliation{Univ. Grenoble Alpes, CEA, Leti, 38000 Grenoble, France}
\author{Pierre-Andr\'{e} Mortemousque}
    \affiliation{Univ. Grenoble Alpes, CNRS, Grenoble INP, Institut N\'eel, 38000 Grenoble, France}
    \affiliation{Univ. Grenoble Alpes, CEA, Leti, 38000 Grenoble, France}
\author{Aymeric Richard}
    \affiliation{Univ. Grenoble Alpes, CNRS, Grenoble INP, Institut N\'eel, 38000 Grenoble, France}
\author{Yuma Okazaki}
    \affiliation{National Institute of Advanced Industrial Science and Technology (AIST), National Metrology Institute of Japan (NMIJ), 1-1-1 Umezono, Tsukuba, Ibaraki 305-8563, Japan}
\author{Shuji Nakamura}
    \affiliation{National Institute of Advanced Industrial Science and Technology (AIST), National Metrology Institute of Japan (NMIJ), 1-1-1 Umezono, Tsukuba, Ibaraki 305-8563, Japan}
\author{Arne Ludwig}
    \affiliation{Lehrstuhl f\"{u}r Angewandte Festk\"{o}rperphysik, Ruhr-Universit\"{a}t Bochum, Universit\"{a}tsstra\ss e 150, 44780 Bochum, Germany}
\author{Andreas D. Wieck}
    \affiliation{Lehrstuhl f\"{u}r Angewandte Festk\"{o}rperphysik, Ruhr-Universit\"{a}t Bochum, Universit\"{a}tsstra\ss e 150, 44780 Bochum, Germany}
\author{Matias Urdampilleta}
    \affiliation{Univ. Grenoble Alpes, CNRS, Grenoble INP, Institut N\'eel, 38000 Grenoble, France}
\author{Tristan Meunier}
    \affiliation{Univ. Grenoble Alpes, CNRS, Grenoble INP, Institut N\'eel, 38000 Grenoble, France}
\author{Tetsuo Kodera}
    \affiliation{Department of Electrical and Electronic Engineering, Tokyo Institute of Technology, Tokyo 152-8550, Japan}
\author{Nobu-Hisa Kaneko}
    \affiliation{National Institute of Advanced Industrial Science and Technology (AIST), National Metrology Institute of Japan (NMIJ), 1-1-1 Umezono, Tsukuba, Ibaraki 305-8563, Japan}
\author{Shintaro Takada}
    \altaffiliation{corresponding authors: 	\href{mailto:shintaro.takada@aist.go.jp;  christopher.bauerle@neel.cnrs.fr}{shintaro.takada@aist.go.jp;\\ christopher.bauerle@neel.cnrs.fr}}
    \affiliation{National Institute of Advanced Industrial Science and Technology (AIST), National Metrology Institute of Japan (NMIJ), 1-1-1 Umezono, Tsukuba, Ibaraki 305-8563, Japan}     
\author{Christopher B\"auerle}
    \altaffiliation{corresponding authors: 	\href{mailto:shintaro.takada@aist.go.jp;  christopher.bauerle@neel.cnrs.fr}{shintaro.takada@aist.go.jp;\\ christopher.bauerle@neel.cnrs.fr}}
    \affiliation{Univ. Grenoble Alpes, CNRS, Grenoble INP, Institut N\'eel, 38000 Grenoble, France}

\date{\today}

\begin{abstract}
The synthesis of single-cycle, compressed optical and microwave pulses sparked novel areas of fundamental research.
In the field of acoustics, however, such a generation has not been introduced yet.
For numerous applications, the large spatial extent of surface acoustic waves (SAW) causes unwanted perturbations and limits the accuracy of physical manipulations.
Particularly, this restriction applies to SAW-driven quantum experiments with single flying electrons, where extra modulation renders the exact position of the transported electron ambiguous and leads to undesired spin mixing.
Here, we address this challenge by demonstrating single-shot chirp synthesis of a strongly compressed acoustic pulse.
Employing this solitary SAW pulse to transport a single electron between distant quantum dots with an efficiency exceeding 99\%, we show that chirp synthesis is competitive with regular transduction approaches.
Performing a time-resolved investigation of the SAW-driven sending process, we outline the potential of the chirped SAW pulse to synchronize single-electron transport from many quantum-dot sources.
By superimposing multiple pulses, we further point out the capability of chirp synthesis to generate arbitrary acoustic waveforms tailorable to a variety of (opto)nanomechanical applications.
Our results shift the paradigm of compressed pulses to the field of acoustic phonons and pave the way for a SAW-driven platform of single-electron transport that is precise, synchronized, and scalable.
\end{abstract}

\maketitle

\section{Introduction}

The generation of a compressed pulse marked a paradigm shift in optics \cite{Strickland1985,Jones2000,Udem2002}, enabling the realization of attosecond experiments \cite{Brabec2000,Cavalieri2007} as well as the synthesis of arbitrary optical waveforms \cite{Shelton2001,Rausch2008} down to the limit of a single cycle of light \cite{Krauss2010}.
Similarly, shaped microwave pulses have been widely used in nuclear magnetic resonance to dynamically control the state of a classical \cite{Freeman1998} or a quantum state \cite{Vandersypen2005}.
Extending this idea to the field of acoustics, only periodic waveforms with arbitrary shapes have been realized so far \cite{Schulein2015}.
However, to synthesize arbitrary phononic waveforms, it is necessary to generate an on-demand acoustic pulse in the single-cycle limit.

Propagating phonons in the form of surface acoustic waves (SAWs) are massively used in the telecommunication industry, and recently, they are finding more and more impressive applications in quantum science \cite{Delsing2019,Gustafsson2014,Satzinger2018,Bauerle2018,Hsiao2020}.
A particularly promising example are SAW-driven quantum experiments in solid-state devices with single flying electrons \cite{Hermelin2011,McNeil2011,Stotz2005,Sanada2013,Bertrand2016,Takada2019,Jadot2021}.
Owing to piezoelectric coupling, a SAW is accompanied by an electric potential that allows single-shot transport of an electron between distant surface-gate-defined quantum dots (QD) \cite{Hermelin2011,McNeil2011}.
The acousto-electric approach allows highly efficient single-electron transfer along coupled quantum rails approaching macroscopic scales \cite{Takada2019} and in-flight preservation of spin entanglement \cite{Jadot2021}.
These properties make SAW-driven electron transport a technique that is promising for proof-of-principle demonstrations of quantum-computing implementations \cite{Barnes2000,Schuetz2015,Bauerle2018}.

However, sound-driven single-electron transport has an intrinsic limitation related to the large spatial extent of the SAW train. 
The quantum state of the flying electron can be disturbed by SAW modulation during the dwell time in the stationary QDs \cite{Bertrand2016}.
Because of the presence of many potential minima accompanying the SAW (typically hundreds) it is furthermore difficult to transport the flying electron with accurate timing.
To overcome the latter problem, a triggered SAW-driven sending process has been developed \cite{Takada2019}.
Requiring one radio-frequency line and one picosecond-voltage-pulse channel per QD, this method is limited to a few electron sources and thus not scalable.
In addition, the triggering technique introduces unwanted electromagnetic crosstalk and potential charge excitation.
Alternatively, replacing the periodic SAW train with a single-cycle acoustic pulse would deliver an elegant sending approach that brings less perturbation and naturally enables synchronized transport from a basically unlimited number of sources.

Here, we present a chirp-synthesis technique to generate on demand a single, strongly compressed acoustic pulse.
To determine the shape of the engineered SAW, we perform time-resolved measurements with a broadband interdigital transducer (IDT) as SAW detector.
By comparison of the experimental data with numerical simulations based on an impulse-response model, we assess the reliability of the synthesis method and outline a path toward maximum pulse compression.
We then employ the acousto-electric chirped pulse to transport a single electron between distant quantum dots and evaluate the transport efficiency.
Triggering the SAW-driven sending process with a picosecond voltage pulse, we then investigate if the electron is fully confined in the central minimum of the chirped pulse.
Finally, we apply a superposition of phase-shifted chirp signals to demonstrate the emission of multiple pulses with precise control on their time delay. 

\section{Pulse compression via chirp synthesis}
A SAW emitted by an IDT is uniquely determined by its electrode design \cite{Morgan,Ekstrom2017,Dumur2019,Lima2003}.
Changing the unit cell pattern allows, for instance, the generation of higher SAW harmonics for the formation of periodic waveforms of arbitrary shapes \cite{Schulein2015}.
The conceptual generalization of this Fourier-synthesis approach is the emission of a \textit{solitary} SAW pulse.
It can be achieved by the so-called chirp IDT whose frequency response is determined by its gradually changing cell periodicity $\lambda_n$.
This nonuniform design has been extensively used in analog electronic filters \cite{Court1969,Atzeni1975} and in radar technologies \cite{Klauder1960}.
In quantum applications, this approach has so far been mainly employed to broaden the IDT's passband \cite{Weiss2018b}.
However, the chirp design can also be employed in an inverse manner---similar to the formation of an ultrashort laser pulse \cite{Strickland1985}---to superpose a quasi-continuum of many elementary SAWs with gradually changing wavelength to a single, distinct, acoustic pulse.

In this work, we aim at the emission of a solitary SAW pulse approaching the form of a Dirac $\delta$ function.
Mathematically, it is approximated via the superposition of a discrete set of frequencies $f_n$,
\begin{equation}
    \delta(t) \propto \int_{-\infty}^{\infty} e^{i\;2\pi\cdot f\cdot t} df \approx \sum_{n=1}^{N} e^{i\;2\pi\cdot f_n\cdot t},
    \label{equ:delta}
\end{equation}
which is mostly destructive, except around the timing $t=0$ where all elementary waves are in phase and thus interfere constructively.

The central idea for synthesizing such a SAW pulse with a chirp-IDT design is to subsequently drive this set of elementary waves with frequencies $f_n$ [see Eq.~(\ref{equ:delta})] according to its gradually changing cell periodicity $\lambda_n=v_{\textrm{SAW}}/f_n$ -- where $v_{\textrm{SAW}}$ indicates the SAW velocity.
Applying an input signal
\begin{equation}
    V_{\rm S}(t) \propto \sin\left(2\pi\int_0^t f(\tau) d\tau+\phi_0 \right),
    \label{equ:input}
\end{equation}
with properly chosen frequency modulation $f(t)$, the chirp transducer allows us to excite the elementary waves with frequency $f_n$ at the right timing to achieve the desired superposition.
Owing to the widely linear SAW dispersion---see Appendix ~\ref{suppl:disp}---the shape of the emitted pulse remains unchanged during propagation.

The design of the chirp IDT is determined by the set of frequencies $f_n$. 
A natural choice for $f_n$ is an evenly spaced set,
\begin{equation}
    f_n = f_1+(n-1)\cdot\Delta f,\hspace{1cm}\Delta f = \frac{f_N-f_1}{N-1},
    \label{equ:evfreq}
\end{equation}
leading to the following recurrence relation for the cell periodicity:
\begin{equation}
    \lambda_{n+1} = \frac{1}{\frac{1}{\lambda_n}+\frac{\Delta f}{v_\textrm{SAW}}}
    \label{equ:recur}
\end{equation}

With this chirp geometry, maximal pulse compression is achieved by applying an input signal---see derivation in Appendix~\ref{suppl:deriv}---with frequency modulation that follows an exponential course:
\begin{equation}
    f(t) = f_1 \cdot e^{\Delta f \cdot t}
    \label{equ:fmod}
\end{equation}

\begin{figure}[b]
\includegraphics[width=\scwidth]{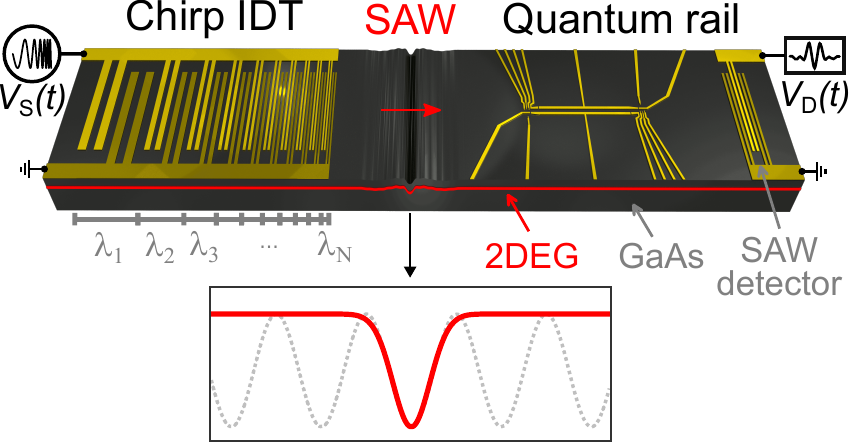}
\caption{Experimental setup.   		
    Schematic of chirp IDT launching a compressed SAW towards a quantum rail and a subsequent broadband SAW detector. 
    We show a perspective view on the sample that is realized via metallic surface gates in a GaAs/AlGaAs heterostructure.
    Inset: comparison between the periodic SAW modulation from regular transduction (dotted gray line) with the ideal SAW profile for electron transport consisting of a single propagating minimum (red line).
    \label{fig:setup}
    }
\end{figure}

\begin{figure*}[t]
\includegraphics[width=\dcwidth]{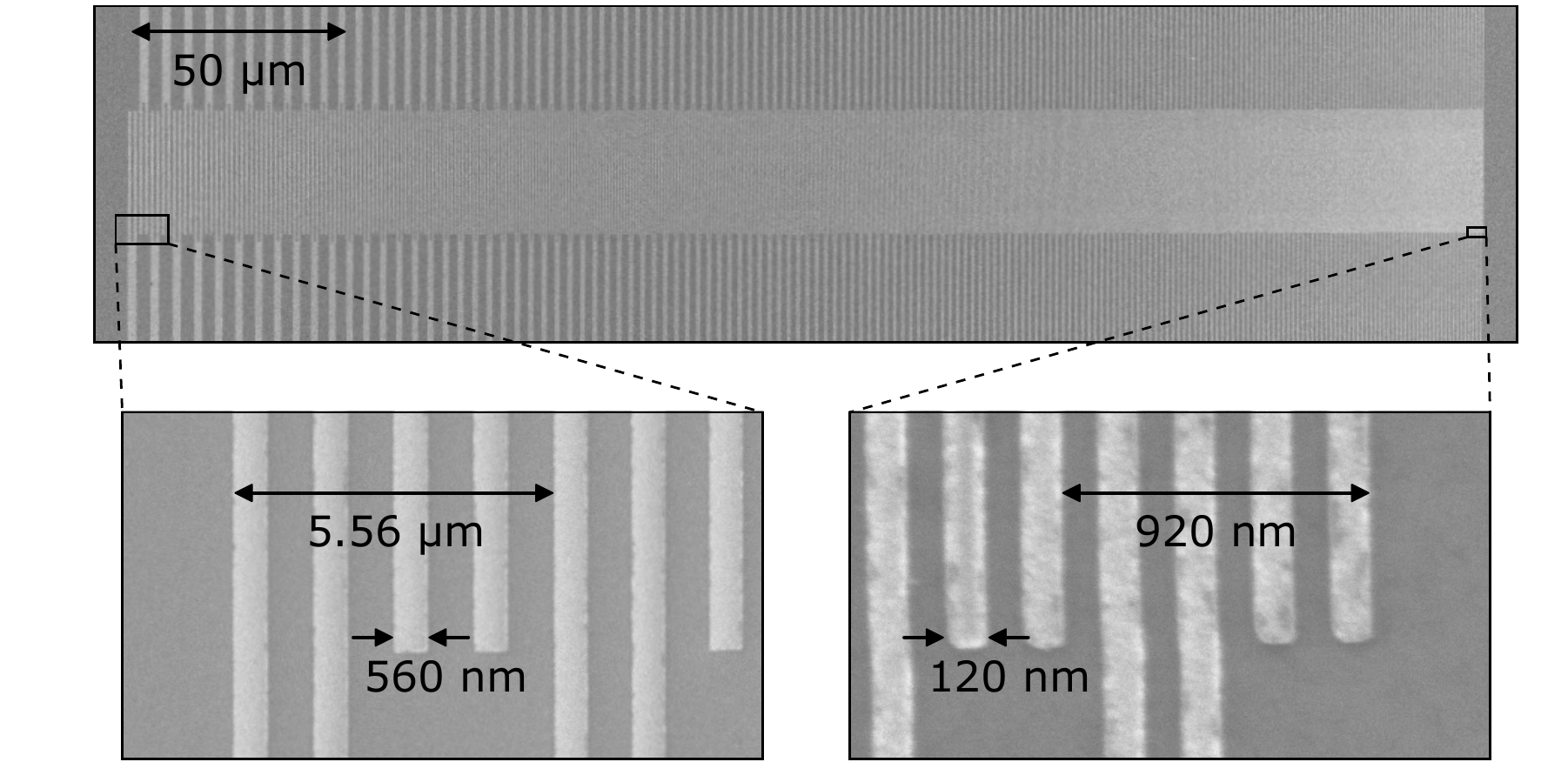}
\caption{SEM images of the chirp IDT with zooms (insets) in the regions of large (left panel) and small (right panel) periodicity of the interlocked electrodes.
    \label{fig:chirp}
    }
\end{figure*}

\section{Experimental setup}

To perform single-electron-transport measurements with the SAW pulse, we employ the experimental setup sketched in Fig.~\ref{fig:setup}. 
The sample consists of a quantum rail that is sandwiched between a chirp IDT and a SAW detector. 
By properly driving the chirp transducer with an input signal $V_{\rm{S}}$, a single propagating SAW minimum is emitted.
When the acoustic pulse passes the quantum rail, the accompanying potential modulation forms a moving QD, which we use to transport an electron in a single shot from one QD to the other. 
The SAW detector that is placed after the quantum rail---see Appendix ~\ref{suppl:detect}---allows us to time-resolve the SAW profile via the induced voltage $V_{\rm D}$, and thus to verify its shape {\it in situ}.

The experiment is performed at a temperature of about 20~mK in a $^3\textrm{He}/^4\textrm{He}$ dilution refrigerator. 
We use a Si-modulation-doped GaAs/AlGaAs heterostructure grown by molecular beam epitaxy (MBE).
The two-dimensional electron gas (2DEG) is located 110~nm below the surface, with an electron density of $n\approx2.8\times 10^{11}$~cm$^{-2}$ and a mobility of $\mu\approx 9\times10^5$~cm$^{2}$V$^{-1}$s$^{-1}$. 
Metallic surface gates (Ti 3~nm, Au 14~nm) define the nanostructures. 
We apply a voltage of 0.3~V on all Schottky gates during cooldown.
At low temperatures, the 2DEG below the transport channel and the QDs is completely depleted via a set of negative voltages applied on the surface gates. 

The surface electrodes of the IDTs are fabricated using standard electron-beam lithography with successive thin-film evaporation (metallization Ti 3~nm, Al 27~nm).
A detailed fabrication recipe is provided in Appendix~\ref{suppl:fab}. 
To reduce internal reflections at resonance, we employ a double-electrode pattern for the transducers. All IDTs have an aperture of 30~\textmu m, with the SAW propagation direction along $[1\bar{1}0]$.
The IDTs are designed and simulated with the homemade open-source Python library ``idtpy" \cite{idtpy}.
We verify the linearity of SAW dispersion in the frequency range of 1 to 8~GHz by employing regular transducers ($\lambda_n=\lambda_0$) on a GaAs substrate (see Appendix~\ref{suppl:disp}). From this investigation, we can deduce the SAW velocity $v_{\rm{SAW}} = (2.81 \pm 0.01)$~\textmu m/ns at ambient temperature.

To characterize the frequency response of the transducer, we measure the transmission $S_{21}$ between two identical IDTs that are opposing each other via a vector network analyzer (Keysight E5080A).
In order to remove parasitic signals from reflections at the sample boundaries, the transmission data are cropped in the time domain after Fourier transform in the range of 300 to 600~ns (expected arrival of first transient around 310~ns) and then transformed back in the frequency domain.

For the time-resolved measurements of the SAW profile, we employ an arbitrary waveform generator (AWG, Keysight M8195A) to provide the input signal $V_{\rm{S}}$ of the chirp IDT.
We record the induced voltage $V_{\rm D}$ on the detector IDT via a fast sampling oscilloscope (Keysight N1094B DCA-M).
In order to reinforce the input and detection signals, $V_{\rm S}$ and $V_{\rm D}$, broadband amplifiers (SHF S126A) are placed along the transmission lines that are connected to the respective IDT's.
As for the transmission data, we apply a Fourier filter on the time-resolved data in the range of 0.4 to 3.5~GHz in order to suppress parasitic contributions from internal higher harmonics of the AWG, the amplifier responses, airborne capacitive coupling, and standing waves in the rf lines. 

\section{Generation of an acoustic chirped pulse}

The synthesis of the strongly compressed acousto-electric pulse is performed with a chirp transducer as shown via the scanning-electron-microscopy (SEM) image in Fig.~\ref{fig:chirp}.
It consists of $N=167$ cells ranging from $f_1 = $ 0.5~GHz to $f_N = $ 3~GHz with the cell periodicity gradually changing from $\lambda_1\approx$ 5.56~\textmu m to $\lambda_N\approx$ 0.92~\textmu m according to Eq.~(\ref{equ:recur}). 

\begin{figure}
\includegraphics[width=\scwidth]{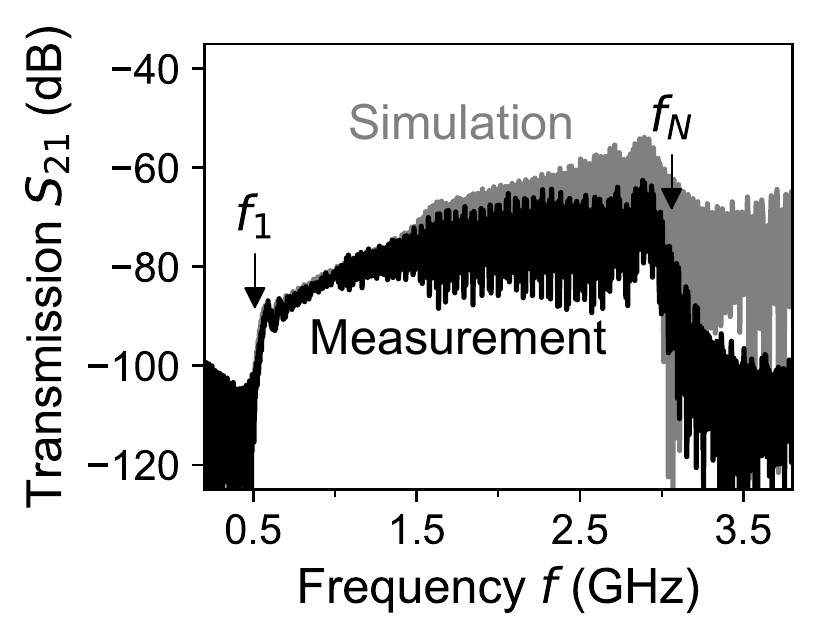}
\caption{Frequency response.		
    Transmission measurement between opposing chirp IDTs (black) with simulation via the delta-function model (gray) and indications of the passband ranging from $f_1\approx$ 0.5~GHz to $f_N\approx$ 3.0~GHz. Note that the transmission bandwidth of the radio-frequency lines is not considered in the simulation.
	\label{fig:vna}
    }
\end{figure}

The transmission spectrum of the chirp IDT allows us to benchmark its quality via the shape of the passband.
Figure~\ref{fig:vna} shows transmission data from a measurement at ambient temperature (black) and the expectation from a delta-function model \cite{Tancrell1971} (gray; see Appendix~\ref{suppl:dfmodel}).
We observe a continuous spectrum over a broad frequency range that is defined by the varying finger periodicity $\lambda_n$.
The flatness of the chirp IDT's passband is mainly achieved by the light electrode material (aluminium) mitigating resonance shifts from mass loading \cite{Morgan}.
The good agreement between experiment and simulation reflects the well-controlled nanofabrication process.

\begin{figure}
\includegraphics[width=\scwidth]{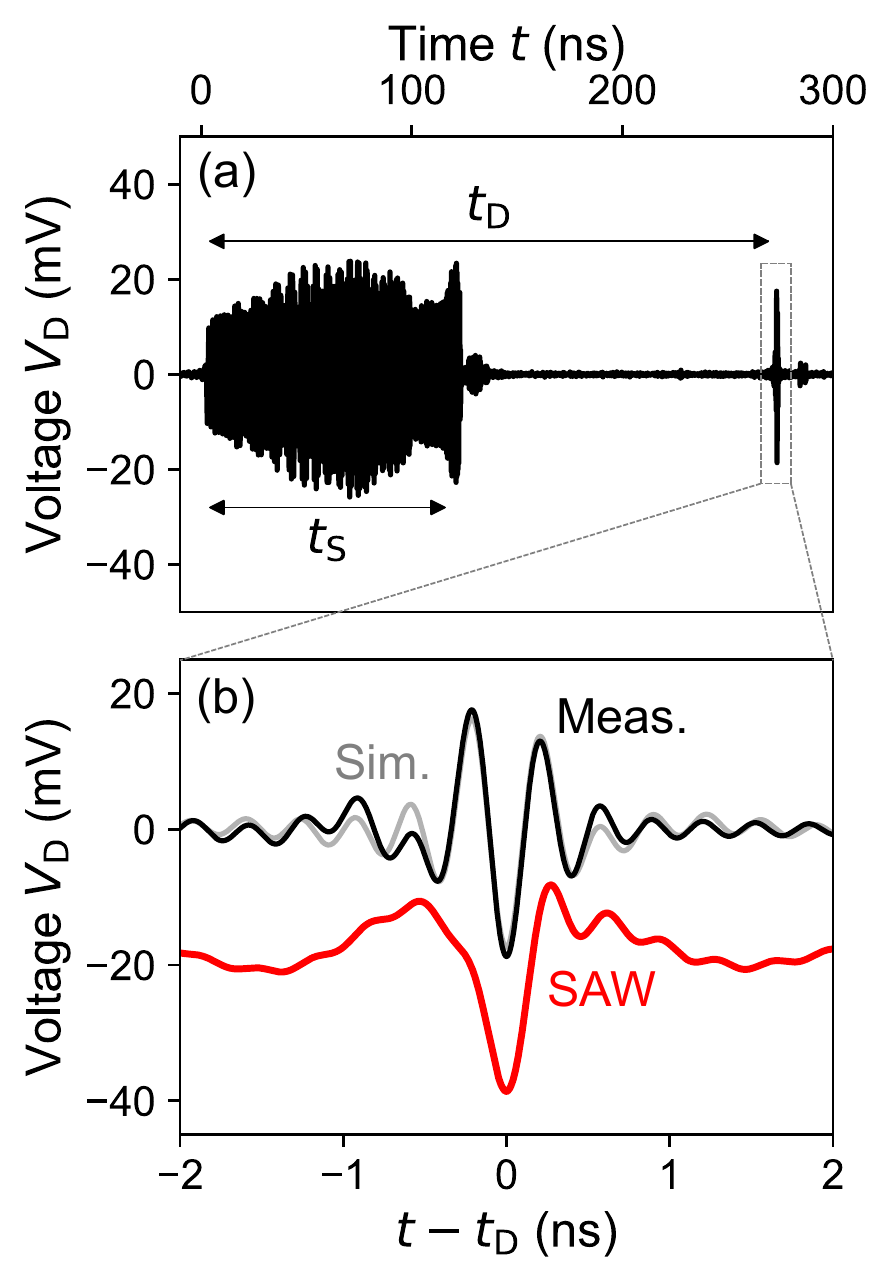}
\caption{Time-resolved measurements.		
    (a)
	Trace of detector response for the frequency-modulated input signal applied on the chirp IDT with $t_\textrm{S}\approx 120$~ns. After initial electromagnetic crosstalk ($t=0$~ns), SAW arrives at the expected delay time $t_\textrm{D}$.
	(b) Zoom-in of the time window around $t_\textrm{D}$, which shows the detected voltage pulse (black) with impulse-response simulation (gray) and the corresponding SAW shape (red; with offset and arbitrary units) derived via deconvolution of the detector response.
	\label{fig:trace}
    }
\end{figure}

Having outlined the basic properties of the chirp IDT, let us now employ it for single-shot pulse generation.
For maximal pulse compression, we apply an input signal that follows Eq.~(\ref{equ:input}) and (\ref{equ:fmod}) with a duration $t_{\rm{S}}$ matching the SAW-propagation time along the transducer, $t_N\approx 120$~ns---see Appendix~\ref{suppl:dev}.
The measured time-resolved response on the SAW detector at room temperature is shown in Fig.~\ref{fig:trace}(a). 
We observe an initial electromagnetic crosstalk (at time $t=0$~ns) followed by a SAW-related response that appears at the expected delay $t_{\rm{D}}$.
The clear contrast between the input signal duration $t_{\rm{S}}$ and the narrow SAW signal confirms the successful compression.
Zooming in on the arrival window [see Fig.~\ref{fig:trace}(b)], we observe the narrow response which follows the shape expected from the impulse-response model \cite{Hartmann1988}  (see Appendix~\ref{suppl:irmodel}).
A slight asymmetry occurs due to a phase offset $\phi_0\approx\pi/2$ introduced by the amplifier.
Because of the agreement between experiment and simulation, we can extract the actual SAW shape (red line with offset) via deconvolution of the detector-response function.
We find that the actual SAW profile has much flatter sidelobes than the signal $V_{\rm D}$ on the broadband detector.

\section{Single-electron transport}

\begin{figure*}
\includegraphics[width=\dcwidth]{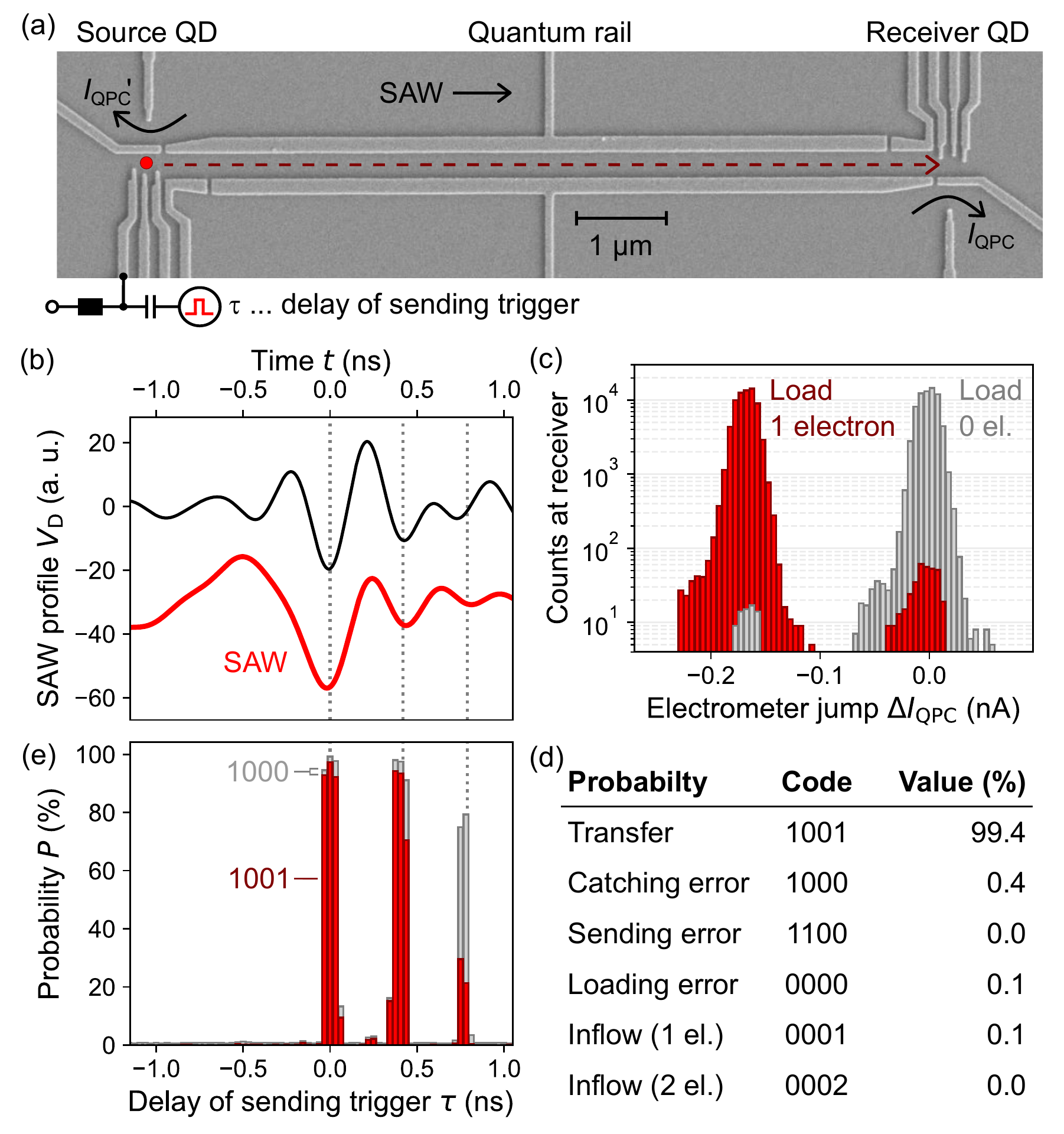}
\caption{Electron shuttling along quantum rail.	
	(a)
	SEM image of the quantum rail consisting of two surface-gate-defined quantum dots (QD) that are connected via a depleted transport channel. 
	We further show the quantum-point-contact (QPC) electrometers that are placed next to each QD to sense the presence of electrons.
	(b)
	Time-dependent measurement of the chirped pulse in the cryogenic setup via the SAW detector (black).
	The offset red line shows the corresponding SAW profile extracted via deconvolution of the detector's response function.
	(c)
    Histogram of jumps in the electrometer current $\Delta I_{\rm QPC}$ at the receiver QD from 70 000 single-shot measurements after launching the SAW pulse with (red) and without (gray) precedent loading of an electron in the source QD.
	(d)
	Table of transfer and error probabilities with QD-occupation code of the events (source before, source after, receiver before, receiver after).
	(e)
	Successful (1001) and failed (1000) transfer probabilities as a function of the trigger delay $\tau$ of the sending process with respect to the SAW emission.
	\label{fig:shuttling}
    }
\end{figure*}

To demonstrate the ability of this highly compressed SAW pulse to transport an electron, let us now employ the 8-µm-long quantum rail.
Figure~\ref{fig:shuttling}(a) shows a SEM image of the nanoscale device.
A QD is located at each end of the transport channel, serving as a single-electron source and receiver.
The occupancy of the QD is monitored by the variation in the current through a nearby QPC acting as a highly sensitive electrometer.
For each transport sequence, we first evacuate all electrons in the system and then load one electron into the source QD (see red point).
Second, the chirp IDT is excited to emit the compressed SAW pulse which then propagates along the quantum rail.
If the SAW is capable of picking up the electron at the source and bringing it to the receiver, related changes are detected in the electron occupancy of the source and receiver QDs.

In order to optimize the SAW profile for single-electron transport in a single moving potential minimum, we exploit the input signal's phase offset $\phi_0$ [see Eq.~(\ref{equ:input})] to form an asymmetric  chirped pulse.
Note that an increased SAW velocity \cite{Powlowski2019} has to be taken into account for the input signal at a cryogenic condition.
Analyzing the SAW profile with $\phi_0\approx 3\pi/2$, we observe a smooth ramp just before the first strongly pronounced minimum ($t<0$~ns) as shown in Fig.~\ref{fig:shuttling}(b).
With this choice, electron transfer is suppressed until the arrival of the leading SAW minimum.
Employing this chirped pulse to perform single-shot electron shuttling with many repetitions, we observe a histogram of QPC-current jumps, $\Delta I_{\rm QPC}$, as shown in Fig.~\ref{fig:shuttling}(c).
As a reference, we also perform each transport sequence without loading an electron at the source QD (gray).
The comparison of the electrometer data at the receiver QD shows sufficient contrast to clearly distinguish transport events.
Moreover, the reference data allows us to quantify the amount of undesired extra electrons injected into the system from outside (inflow).
Figure~\ref{fig:shuttling}(d) summarizes the transfer probability and the sources of error (loading, sending, catching and inflow) from 70 000 single-shot measurements.
The overall low error rates indicate a single-electron-transfer efficiency of $(99.4\pm0.4)$\%, which is similar to the highest value achieved with regular IDT design \cite{Takada2019}.

Let us now focus on the question of where exactly within the compressed SAW pulse the electron is transferred.
For this purpose, we employ a fast voltage pulse injected via a bias tee on a gate of the source QD [see Fig.~\ref{fig:shuttling}(a)] to trigger the sending process with the SAW \cite{Takada2019}.
In this experiment, the potential landscape of the source QD is set such that the initially loaded electron is protected when the acoustic wave passes.
By triggering a picosecond voltage pulse, the potential is temporarily lifted to load the electron into the moving SAW.
Sweeping the time delay $\tau$ of this trigger, we thus successively address each position along the SAW pulse in an attempt to transfer the electron.
Figure~\ref{fig:shuttling}(e) shows transmission probability data of such a measurement.
We observe three transmission peaks that emerge in congruence with the potential minima of the SAW profile.
The highest transport probability (code 1001) appears at the first peak ($\tau=0$~ns) that corresponds to the deepest minimum of the SAW pulse.
The extent of 97\% sets a lower limit to the probability that the electron is emitted on arrival of this moving potential minimum at the source QD.

In order to investigate whether the electron stays within this position as it propagates along the quantum rail, it is insightful to also look at the unsuccessful transfer events (code 1000).
The strongly increased error at the third peak of more than 40\% indicates that it plays a rather negligible role since, without a sending trigger, this error is only 0.4\%.
Estimating an amplitude of $(19\pm3)$~meV of the first acoustic minimum---see Appendix\,\hyperref[suppl:comp]{H}---the currently employed SAW confinement is slightly below the 95\%-confinement threshold of approximately $24$~meV \cite{Edlbauer2021}.
Therefore, we cannot exclude transitions into the second minimum ($\tau\approx0.4$~ns) during transport.
However, we anticipate reinforcement of single-minimum confinement via increased input-signal power and enhanced transducer design.
We further evaluate the orbital level spacing by approximating the acoustic minimum to a parabolic potential \cite{Ciftja2009}.
For instance, if the frequency range is raised up to 6~GHz, we find an increase of the energy spacing from 2~meV to 3~meV.
Accordingly, we expect that the reinforcement of the SAW confinement will also enable the loading of a single electron into the ground state and transport without excitation \cite{Kataoka2009,Takada2019,Ito2021}---two conditions that are essential for the realization of SAW-driven flying electron qubits.

\section{SAW engineering}

\begin{figure}[h]
\includegraphics[width=\scwidth]{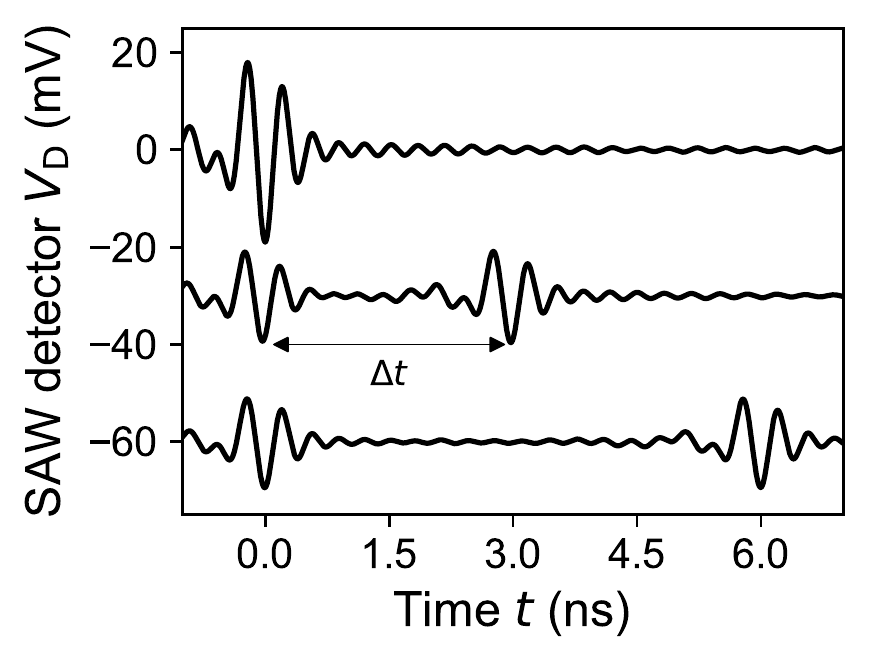}
\caption{Acousto-electric wave engineering.
	Time-resolved measurements from the SAW detector for an input signal consisting of two superposed pulses with different time delays $\Delta t=\Delta t_2-\Delta t_1\in[0,3,6]$~ns. 
	\label{fig:engineering}
    }
\end{figure}

The wide-ranging linearity of the SAW dispersion opens up a flexible platform to engineer any nanomechanical waveform using a single chirp IDT.
Multiple $\delta$ pulses can be superposed via overlaid input signals $V_p(t)$ with deliberately chosen delay ($\Delta t_p$), phase ($\phi_p$), and amplitude ($A_p$):
\begin{equation}
    V(t) = \sum_{p=1}^P A_p \cdot V_p(t+\Delta t_p, \phi_p)
\end{equation}
Following this approach, a sawtooth shape can be achieved, for instance, by superimposing uniformly delayed pulses with linearly decreasing amplitude.
For the sake of simplicity, let us demonstrate this wave-engineering method by means of two pulses ($P=2$) with arbitrary delay.
A relevant application of such a synthesis is the sequential transport of a pair of entangled electrons to observe spin interference patterns \cite{Jadot2021}.
Figure~\ref{fig:engineering} shows the SAW profile from time-dependent measurements of such an engineered waveform.
Two identical pulses are apparent, and they are separated by the chosen delay $\Delta t$.
Note that the halving in pulse amplitude compared to the single-pulse case ($\Delta t=0$) is expected since the amplitude scales inversely with the number of superposed signals $P$ (for $\Delta t<t_N$).
In order to achieve sufficient amplitude of the engineered waveform, it is thus crucial to maximize the IDT length (via the number of cells $N$) and the bandwidth ($B=f_N-f_1$) to achieve maximal pulse compression---see Appendix~\ref{suppl:rules}.
Owing to the linear SAW dispersion, the shape of the generated pulses is independent of the delay of the input signals $V_1$ and $V_2$.
The precise time control of $\delta$ pulses lays the groundwork for on-demand emissions of arbitrary nanomechanical waveforms.

\section{Summary and outlook}

In conclusion, we have demonstrated an original SAW-engineering method to generate in a single shot a solitary acousto-electric $\delta$ pulse.
We implemented the concept using a chirp transducer operating in the frequency band of 0.5 to 3~GHz.
Our investigations showed that chirp synthesis is a highly controlled technique allowing reliable acoustic pulse shaping by design.
Demonstrating a single-electron transport efficiency exceeding 99\%, we confirmed robust potential confinement for SAW-driven quantum transport.
Confirming the confinement location during flight, this acoustic chirped wavefront thus represents the scalable alternative for synchronized and unambiguous SAW-driven single-electron transport from multiple sources.
This technique is compatible with all the essential building blocks developed for SAW-driven flying electron qubits such as on-demand single-electron partitioning \cite{Takada2019}, time-of-flight measurements \cite{Edlbauer2021} and electron-spin transfer \cite{Bertrand2016, Jadot2021}.
The nonuniform IDT design enables the possibility to engineer arbitrary combinations of superposed pulses having high relevance for experiments where multiple charges are transferred successively \cite{Jadot2021}.
Accordingly, we expect that the chirp approach opens up new routes for quantum experiments on interference and entanglement exploiting spin and charge degree of freedom with single flying electrons \cite{Bauerle2018,Barnes2000,Schuetz2015}.

We highlighted that chirp synthesis is readily applicable to other piezoelectric platforms such as LiNbO$_3$ or ZnO.
Further enhancement of the power density is achievable by integrating the concept of unidirectional \cite{Ekstrom2017,Dumur2019} or focusing \cite{Lima2003} designs in the chirp transducer.
Additionally, the frequency band and thus pulse compression is easily adjustable via the electrode periodicity of the chirp IDT.

Finally, owing to the wide-ranging applications of propagating phonons in fundamental research \cite{Delsing2019,Gustafsson2014,Satzinger2018,Bauerle2018,Hsiao2020,Midolo2018,Yokoi2020,Kobayashi2017,Yokouchi2020,Chen2021}, the demonstrated acoustic pulse is not restricted to the field of quantum information processing.
In spintronics, for instance, our chirp synthesis technique opens up the way for on-demand generation of spin-current pulses in nonferromagnetic materials \cite{Kobayashi2017}.
Employing two compressed acoustic pulses with a controlled time delay and opposite phases, it further allows investigations on the spin-current formation in the time domain.
Moreover, since SAW can create skyrmions in thin-film samples without Joule heating \cite{Yokouchi2020}, a solitary acoustic pulse could be the key to create a single skyrmion and to perform manipulations at the single-shot level.
In metrology applications, the accuracy of SAW-driven electron pumps \cite{Cunningham2000} is currently limited by the overlapping between the electromagnetic crosstalk and the acoustic signal \cite{Kataoka2006}.
Emitting compressed pulses with a controlled repetition rate, such single-electron pumps can be significantly enhanced in performance and easily operated in parallel without additional radio-frequency lines.
Similarly, phonons can also stimulate single-photon emission in a hybrid quantum-dot--nanocavity system \cite{Weiss2016}.
Since each SAW period contributes to the creation of photons, our technique would allow on-demand single-photon emission with precise timing.
In summary, analogous to the advantages of using solitary optical \cite{Strickland1985,Jones2000,Udem2002,Brabec2000,Cavalieri2007,Shelton2001,Rausch2008,Krauss2010} and microwave pulses \cite{Freeman1998,Vandersypen2005}, we anticipate that the presented compression technique will open new routes for fundamental research employing nanoscale acoustics.

\begin{acknowledgments}

J.W. acknowledges the European Union's Horizon 2020 research and innovation program under the Marie Skłodowska-Curie Grant Agreement No. 754303.
A.R. acknowledges financial support from ANR-21-CMAQ-0003, France 2030, Projet QuantForm-UGA.
T.K. and S.T. acknowledge financial support from JSPS KAKENHI Grant No. 20H02559.
C.B. acknowledges funding from the European Union’s H2020 research and innovation program under Grant Agreement No. 862683 and from the French Agence Nationale de la Recherche (ANR), project QUABS ANR-21-CE47-0013-01. 
A.D.W., and A.L. acknowledge support from TRR 160/2-Project B04, DFG 383065199, the German Federal Ministry of Education and Research via QR.X Project 16KISQ009, and the DFH/UFA CDFA-05-06.
\end{acknowledgments}

\appendix

\section{LINEARITY OF SAW DISPERSION
\label{suppl:disp}}

\begin{figure}[b]
\includegraphics[width=\scwidth]{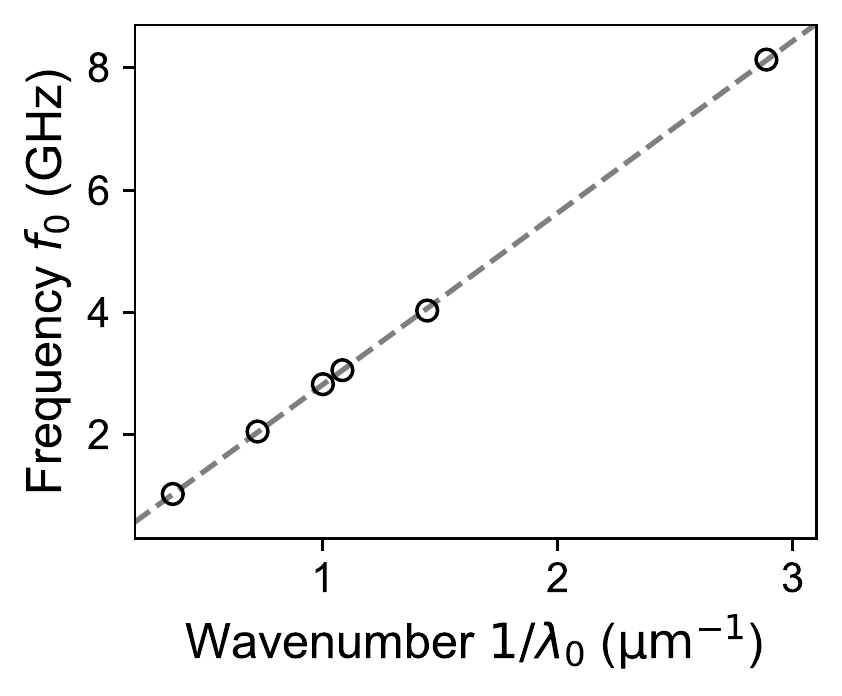}
\caption{SAW-dispersion relation for aluminium IDTs on a GaAs substrate.
	We show the resonance frequency for IDTs having different periodicity---here expressed as wave number $1/\lambda_0=k/2\pi$.
	The points are extracted from Gaussian fits of distinct peaks in the transmission data (reflection data for the case of 8~GHz) from a network analyzer measurement.
	\label{sfig:disp}
    }
\end{figure}

To test the linearity of the SAW dispersion for the frequency range of interest, we investigate the resonant response of six regular aluminium IDTs on a GaAs substrate in transmission measurements at room temperature.
As a starting point for the IDT designs, we consider the SAW velocity $v_\textrm{SAW}^\textrm{Au}\approx2.77$~\textmu m/ns for gold transducers, WHICH is well known from former SAW-driven charge transport experiments.
For the resonance frequencies, we target $f_0\in[1,2,2.77,3,4,8]$~GHz, giving a periodicity of the respective IDTs of $\lambda_0 = v_\textrm{SAW}^\textrm{Au}/f_0$.
Because of the reduced mass loading of aluminium electrodes, we expect a slightly increased resonance frequency of the respective transducers.
Figure~\ref{sfig:disp} shows a plot of the resonance frequencies of the fabricated IDTs as a function of the wave number $k=2\pi/\lambda_0$.
The data show a linear dispersion in the investigated frequency range of 1 to 8~GHz.
Performing a linear least-squares fit of the data (see dashed line), we can deduce the SAW velocity for aluminium IDTs on GaAs via its slope as $v_\textrm{SAW}= (2.81 \pm 0.01)$~\textmu m/ns.

\section{DERIVATION OF EXPONENTIAL FREQUENCY MODULATION
\label{suppl:deriv}}

In order to have in-phase interference of the $N$ elementary waves at the IDT boundary, the frequency modulation $f(t)$ of the input signal must introduce the $n$th elementary SAW with the right delay, $t_n$.
For an evenly spaced set of frequencies [Eq.~(\ref{equ:evfreq})] with a single period per step $n$, we obtain the excitation times:
\begin{equation}
    t_n = \sum_{m=1}^{n} \frac{1}{f_m}.
    \label{equ:extime}
\end{equation}
The resulting sum can then be expressed in terms of the digamma function $\Psi$:
\begin{equation}
    t_n = \sum_{m=1}^{n} \frac{1}{f_1+(m-1)\cdot \Delta f} = \frac{\Psi \left( \frac{f_1}{\Delta f}+n \right) - \Psi \left( \frac{f_1}{\Delta f} \right)}{\Delta f}.
    \label{sequ:extime}
\end{equation}
Since $f_1/\Delta f \gg 1$, this function approaches a logarithmic course:
\begin{equation}
    \Psi(x) \overset{x \gg 1}{\approx} \ln(x).
\end{equation}
Multiplying Eq.~(\ref{sequ:extime}) by $\Delta f$ and additionally applying an exponential function, we thus obtain
\begin{equation}
    e^{\Delta f\cdot t_n} = 
    (\underbrace{f_1+(n-1)\cdot\Delta f}_{f_n}+\Delta f)/f_1
\end{equation}
Which brings us to the desired frequency modulation of the input signal:
\begin{equation}
    f_n = f_1\cdot e^{\Delta f\cdot t_n} - \Delta f \overset{f_1 \gg \Delta f}{\approx} f_1\cdot e^{\Delta f\cdot t_n}
\end{equation}
as expressed in its continuous form $f(t)$ in Eq.~(\ref{equ:fmod}).

\section{BROADBAND DETECTION
\label{suppl:detect}}

Owing to the inverse piezoelectric effect, when a SAW passes through another IDT, an electric signal is induced that can be recorded by a fast sampling oscilloscope.
To optimize the response of the detector IDT, it is necessary to design the transducer according to the expected bandwidth of the input signal.

\begin{figure}
\includegraphics[width=\scwidth]{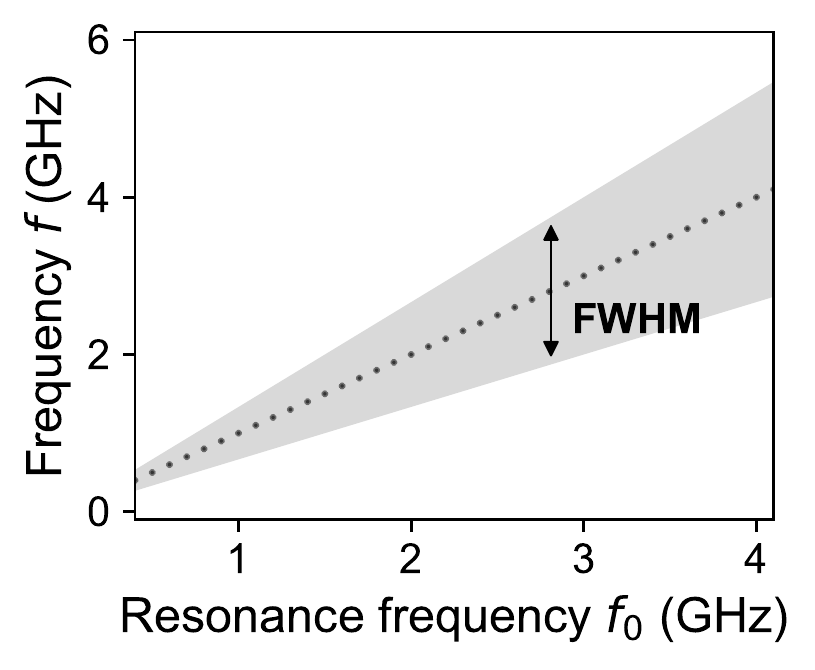}
\caption{Detection bandwidth.
	Pass band of a regular IDT with $N=1.5$.
	The shaded area indicates the FWHM of the transmission peak.
	The vertical double-headed arrow indicates the employed resonance frequency of the detector IDT, $f_0\approx2.81$~GHz.
	\label{sfig:det}
    }
\end{figure}

Generally, the frequency response of a regular IDT with $N$ unit cells and resonant frequency $f_0$ follows approximately a sinc function \cite{Morgan}:
\begin{equation}
     H(f) \propto \frac{\sin{(\pi N (f-f_0)/f_0)}}{\pi N (f-f_0)/f_0}
\end{equation}
Then the full width at half maximum (FWHM) can be derived as $f_0/N$.
Because of the large bandwidth of the chirped pulse, we minimize the number of detector electrodes to one, giving $N = 1.5$ with a pair of neighboring grounded fingers.
Figure~\ref{sfig:det} shows the passband for this geometry as a function of $f_0$.
In order to reliably resolve the SAW response particularly at 3 GHz and below, we have chosen a detector periodicity of $\lambda_0 =1$~\textmu m ($f_0\approx2.81$~GHz).

\section{NANOFABRICATION RECIPE
\label{suppl:fab}}

The IDTs were fabricated according to the steps listed in Table~\ref{tab:fab}.
A metallization ratio of 0.4--0.5 was chosen for all of the interlocked transducer fingers.
At electron-beam lithography, a proximity correction was applied for all IDTs.

\begin{table*}
\caption{Fabrication procedure} \label{tab:fab}
\begin{ruledtabular}
\begin{tabular}{lll}
Number&Method&Description\\ \hline
1 & Ultrasonic cleaning & 10~min in acetone + 10~min in isopropanol (IPA)\\ 
2 & Warming on hotplate & 115$\rm\;{^{\circ}C}$ for 2~min + wait 5~min to cool down \\ 
3 & Spin coating of resist & PMMA 3\%, 4000~rpm, 4000~rpm/s, 60~s + bake at 180$\rm\;{^{\circ}C}$ for 5~min \\
4 & Electron-beam lithography & Writing of IDT structures. Equipment: \textit{Nanobeam nB5} \\
5 & Development of resist & 35~s in MIBK:IPA 1:3 + 1~min in IPA \\
6 & Oxygen-plasma cleaning & With a power of 10~W for a duration of 10~s \\
7 & Metal deposition & Ti 3~nm at 0.05~nm/s + Al 27~nm at 0.10~nm/s \\
8 & Lift-off & N-methyl-2-pyrrolidone (NMP) at 80$\rm\;{^{\circ}C}$ for at least 1~h\\ 
9 & Ultrasonic cleaning & 20~s in acetone + 20~s in IPA\\ 
10 & Spin coating of resist & S1805, 6000~rpm, 6000~rpm/s, 30~s + bake at 115$\rm\;{^{\circ}C}$ for 1~min\\ 
11 & Laser lithography & Writing of contacts and ground plane\\ 
12 & Development of resist & 1~min in Microposit developer:DI H$_2$O 1:1 + 1~min in DI H$_2$O \\ 
13 & Metal deposition & Ti 20~nm at 0.10~nm/s + Au 80~nm at 0.15~nm/s \\
14 & Lift-off & At least 30~min in acetone \\ 
\end{tabular}
\end{ruledtabular}
\end{table*}

\section{DELTA-FUNCTION MODEL
\label{suppl:dfmodel}}

The simplest way of modeling the IDT response is to approximate the output as a superposition of elementary waves that are emitted with delay times $t_n$ at discrete point sources located at the finger positions $x_n$.
In this picture, the response function for a certain IDT geometry can be written in the time domain as a sum of Dirac delta functions located at each finger location:
\begin{equation}
    h(t) = \sum_{n=0}^{N} P_n \cdot \delta(t-t_n)\label{equ:delta2}
\end{equation}
where $P_n\in\pm 1$ is the polarity of the $n$th finger that indicates connection to the input electrode or to the ground.
In general, the SAW response $y(t)$ can be mathematically expressed as a convolution of an input signal $V(t)$ with the IDT geometry $h(t)$ acting as a filter:
\begin{equation}
    y(t) = (V\ast h) (t).
    \label{equ:yconv}
\end{equation}
The IDT response in the frequency domain can then be calculated by the Fourier transform of $y(t)$ via application of the convolution theorem as
\begin{equation}
    \hat{y}(\omega) = \hat{V}(\omega)\cdot \hat{h}(\omega),
\end{equation}
where $\hat{V}$ and $\hat{h}$ indicate, respectively, the Fourier transform of $V$ and $h$.
Considering a continuous input signal, $V(t)\propto e^{i\;\omega_0\cdot t}$, we obtain
\begin{equation}
    \hat{V}(\omega) = \int_{-\infty}^\infty e^{i\;(\omega_0-\omega)\cdot t} dt \propto \delta(\omega_0-\omega),
\end{equation}
Which brings us to
\begin{equation}
    \hat{y}(\omega) \propto \hat{h}(\omega).
\end{equation}
To obtain the frequency response of a certain IDT geometry within the delta-function model, we can thus directly evaluate the Fourier transform of Eq.~(\ref{equ:delta}):
\begin{align}
    \hat{y}(\omega) \propto \sum_{n=0}^{N} P_n \int_{-\infty}^\infty \delta(t-t_n) \cdot e^{i\;\omega\cdot t} dt = \sum_{n=0}^{N} P_n\cdot e^{i\;\omega\cdot t_n},
\end{align}
where $t_n=x_n/v_\textrm{SAW}$ is determined by the SAW velocity $v_\textrm{SAW}$ and the IDT's finger positions $x_n$.

\section{COMPENSATION WITH INPUT SIGNAL FOR PULSE COMPRESSION
\label{suppl:dev}}

\begin{figure}[!h]
\includegraphics[width=\scwidth]{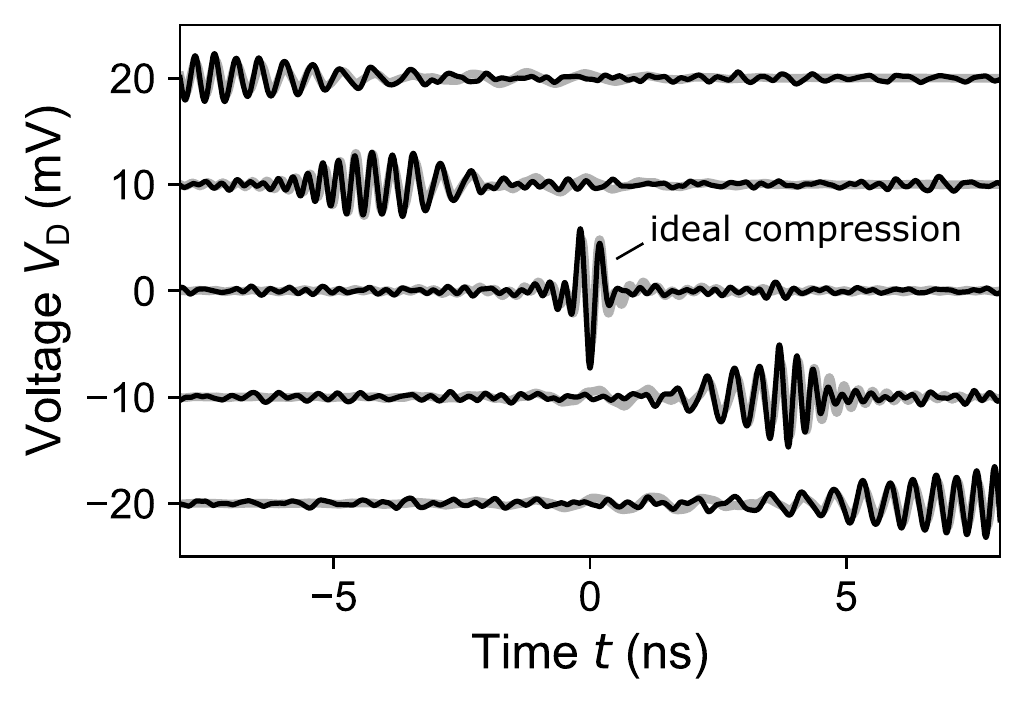}
\caption{Nonideal input signals.
	Time-resolved measurements of the SAW response for an input signal with duration deviation of $t_\textrm{S} - t_{N}\in[-10, -5, 0, 5, 10]$~ns.
	The semitransparent line in the background shows the course expected from an impulse-response model.
	\label{sfig:dev}
    }
\end{figure}

For optimum acoustic chirped pulse compression, it is important that the input signal matches the IDT response, $V_{\rm S}(t) = h(t)$.
Typically, however, the fabricated IDTs show slight deviations from the design, which leads to minor changes in the IDT response.
To compensate this irregularity for maximal pulse compression, we exploit the input signal parameters such as $t_{\rm{S}}$ or $f_1$ [see Eq.~(\ref{equ:fmod})]. 
Figure~\ref{sfig:dev} shows the broadening effect for input signals with duration $t_\textrm{S}$ deviating from the ideal parameter, $t_N$.
When employing a nonideal input signal [$V_{\rm{S}}(t) \neq h(t)$], each frequency is excited with a certain phase delay between them, resulting in broadening of the SAW pulse. 
Note that the impulse-response model is also able to predict the influence of nonideal input signals. 
Owing to this compensation method, chirp synthesis becomes more robust against small variations due to the limits of precision in IDT fabrication.

\section{IMPULSE RESPONSE-MODEL
\label{suppl:irmodel}}

\begin{figure}[!b]
\includegraphics[width=\scwidth]{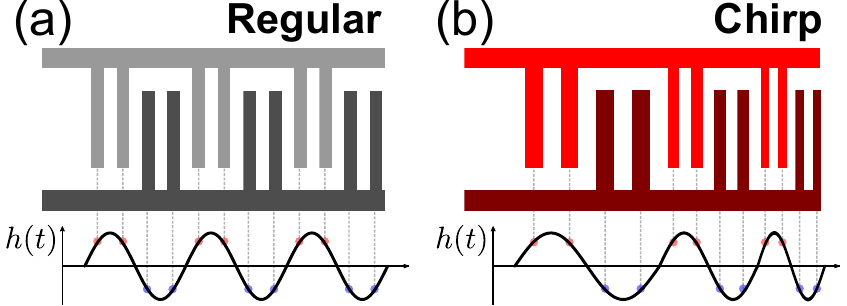}
\caption{Construction of the impulse response of IDTs.
    We show examples of the response function $h(t)$ (vertical axis) in comparison to the finger positions for 
    (a)
    a regular IDT and
    (b)
    a chirp IDT.
    \label{sfig:idt}
    }
\end{figure}

A more accurate description of the IDT geometry can be achieved by the so-called impulse-response model.
Such precision is particularly important when describing the time response of irregular IDT geometries.
In contrast to the aforementioned delta-function model, the time response of an IDT, $h(t)$, is defined by a continuous frequency-modulated function:
\begin{equation}\label{sequ:ht}
h(t) \propto \sin\left(2\pi \int_0^t f(\tau) d\tau\right).
\end{equation}
To construct $h(t)$ via the instantaneous frequency response $f(t)$, one-half cycle of a sine wave is placed between two electrodes with opposite polarity (half period).
This construction rule is schematically shown in Fig.~\ref{sfig:idt} via the examples of a regular and a chirp IDT. 
For a regular IDT, each period is the same, giving a constant frequency response $f(t)=f_0$.
On the other hand, for an irregular IDT geometry, $f(t)$ changes depending on the finger positions.
In the case of a chirp IDT, its response function $h(t)$ is identical to the ideal input signal $V_{\rm S}(t)$ introduced in Eq.~(\ref{equ:input}) and (\ref{equ:fmod}).

In order to calculate the surface displacement by a SAW, $y(t)$, resulting from a certain input signal $V_{\rm S}(t)$, one can now simply calculate the convolution with the IDT response $h(t)$---compare Eq.~(\ref{equ:yconv}).
However, the experimental data that is obtained in a time-resolved measurement also contains information of the detector IDT response, $h_\textrm{D}(t)$.
To compare the time-resolved data to a simulation of an impulse-response model, it is necessary to also convolve $y(t)$ with the detection filter $h_\textrm{D}(t)$:
\begin{equation} \label{sequ:dconv}
V_\textrm{D}(t) \propto (y\ast h_\textrm{D})(t) = (V_\textrm{S}\ast h\ast h_\textrm{D})(t)
\end{equation}

The amplitude of the piezoelectrically introduced voltage $V_\textrm{D}$ depends on several parameters such as the efficiency of the electromechanical coupling or the circuit impedance network.
In this work, we only take into account the geometries of the input IDT and detector IDT, $h$ and $h_\textrm{D}$, and the input signal $V_{\rm S}(t)$.
In order to reproduce the experimental data, we use the amplitude and the SAW velocity as fitting parameters.
The SAW shapes shown in Figs.~\ref{fig:trace}(b) and \ref{fig:shuttling}(b) (red and offset) correspond to the actual SAW profile $y(t)$, where the detector's response is not taken into account.

\section{AMPLITUDE COMPARISON OF CHIRPED PULSE TO SAW STEMMING FROM REGULAR TRANSDUCER
\label{suppl:comp}}

\begin{figure}[!b]
\includegraphics[width=\scwidth]{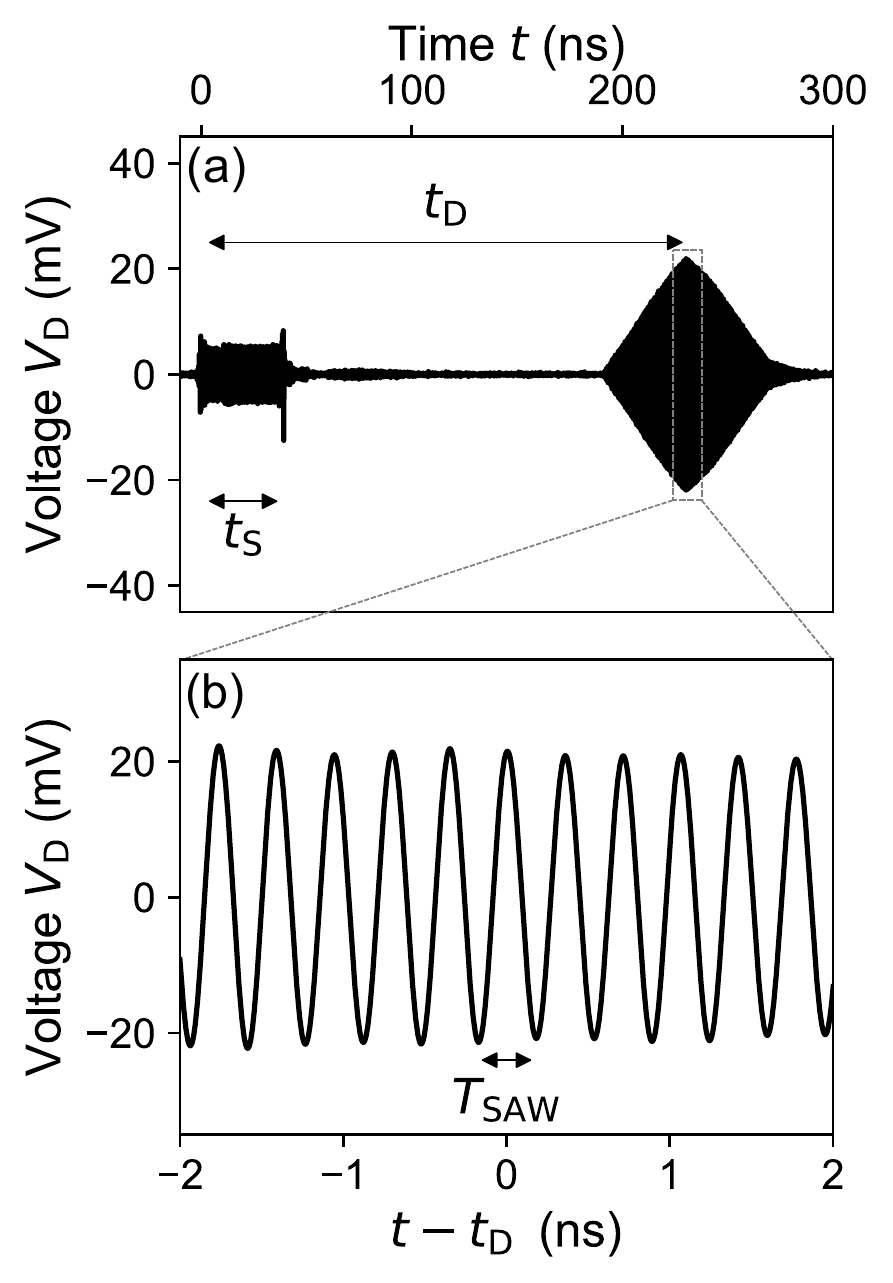}
\caption{Time-resolved measurements on regular IDT.
	(a)
    Trace of detector response for resonant input signal that is applied for a duration $t_\textrm{S}=40$~ns on the regular IDT.
    The SAW arrives with expected delay $t_\textrm{D}$ after initial electromagnetic crosstalk.
    (b) Zoom in SAW region at $t_\textrm{D}$.
	\label{sfig:trm_reg}
    }
\end{figure}

\begin{figure*}
\includegraphics[width=\dcwidth]{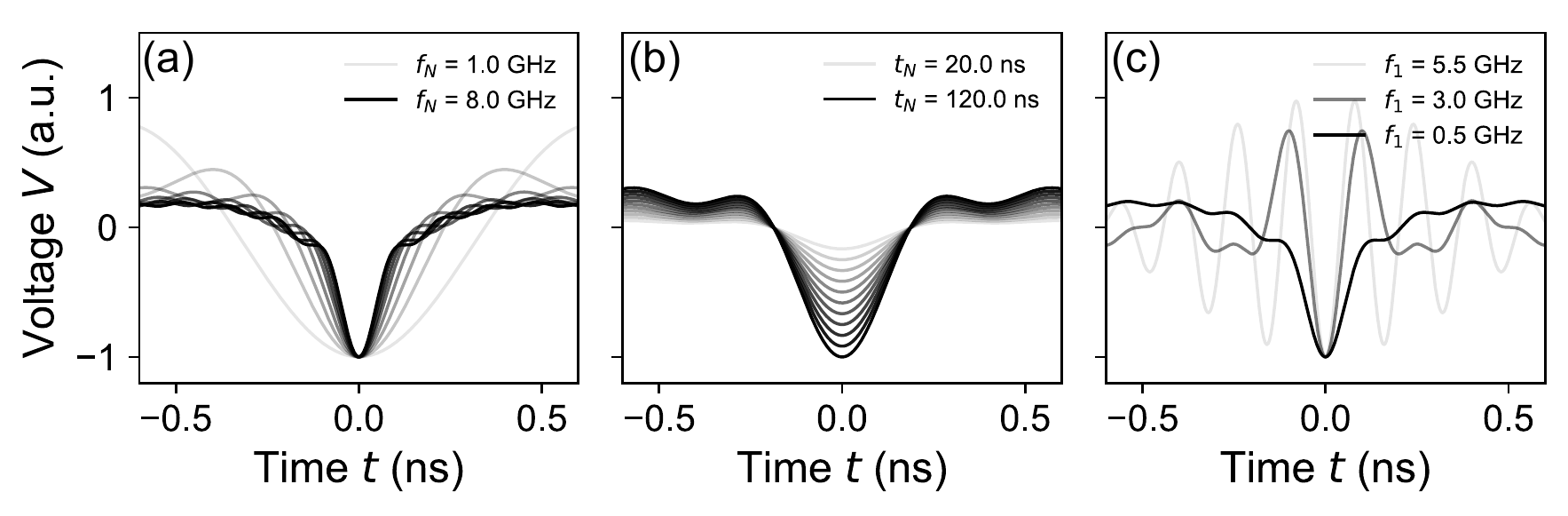}
\caption{Effect of design parameters.
	SAW shapes of impulse-response model for changing IDT parameters.
	(a)
	Varying maximal frequency $f_N$ with $f_1=0.5$~GHz and $t_N=40$~ns.
    (b)
	Increasing IDT length $t_N$ with $f_1=0.5$~GHz and $f_N=3.5$~GHz.
	(c)
	Reducing minimum frequency $f_0$ with $f_N=7$~GHz and $t_N=40$~ns.
	\label{sfig:dp}
    }
\end{figure*}

Let us estimate the amplitude of the chirped pulse employed in this work with the SAW stemming from the regular IDT employed in the time-of-flight measurements reported in a previous work \cite{Edlbauer2021}.
Such a comparison is valid since the chirped pulse experiment was conducted under the same experimental conditions of the flight-time measurement---same fabrication and measurement setup.
The regular IDT consists of $N=111$ cells of period $\lambda_0=1$~\textmu m. 
The resonance frequency that we expect for this reference IDT is $f_0=v_{\rm{SAW}}/\lambda_0\approx2.81$~GHz.
Figure~\ref{sfig:trm_reg} shows time-dependent measurements of a SAW train emitted from the regular transducer with a resonant input signal of duration $t_{\rm{S}}\approx t_{N}$.
This measurement is executed under the same conditions at ambient temperature as the chirp synthesis shown in Fig.~\ref{fig:trace}.
The data show that the chirp signal reaches approximately 80\% of the signal stemming from the SAW train of the regular IDT.
Comparing to the input power of 24.61~dBm sent from ambient temperature to the cryogenic setup with the chirp IDT to the power-to-energy conversion performed in the time-of-flight measurements, we estimate an amplitude of $(19\pm3)$~meV for the compressed SAW pulse.

\section{ANALYSIS OF DESIGN PARAMETERS
\label{suppl:rules}}

The agreement between a time-dependent SAW experiment and simulation enables us to predict the shape of the acoustic pulse for different design parameters of the chirp IDT.
Figure \ref{sfig:dp} shows the evolution of the pulse shape for variations of the maximal frequency $f_N$, the IDT length $t_N$, and the minimum frequency $f_1$.
The simulations allow us to formulate the following design rules for acousto-electric pulse generation with a chirp IDT:
\begin{enumerate}
    \item The pulse narrows with increasing $f_N$ [see Fig.~\ref{sfig:dp}(a)].
    \item The amplitude scales with the number of unit cells $N$ [see Fig.~\ref{sfig:dp}(b)].
    \item Side lobes can be mitigated by reducing $f_1$ [see Fig.~\ref{sfig:dp}(c)]
\end{enumerate}

To form a clean acousto-electric pulse with strong potential confinement, it is thus important to design a chirp IDT with a maximized length and frequency span.
In this regard, we suspect that a single-finger design could be superior to the double-finger pattern employed in this work, since the gain in pulse compression will likely outweigh losses from internal reflections.

\newpage

\bibliography{references.bib}

\end{document}